\documentclass[12pt]{article}
\usepackage{amsthm,amsmath,latexsym,amssymb,amsfonts,amscd}
\usepackage{graphics,lscape,fancyhdr,array,stmaryrd,euscript,wrapfig}
\usepackage{ifthen,empheq,slashed}
\usepackage{ifpdf}
\usepackage{verbatim,slashed}
\numberwithin{equation}{section}
\usepackage{hyperref,setspace}

\usepackage{mathrsfs}
\usepackage[numbers,sort&compress]{natbib}
\usepackage[nottoc]{tocbibind}

\newcommand{\pl}{\partial}

\newcommand{\be}{\begin{equation}}
\newcommand{\ee}{\end{equation}}

\newcommand{\mm}{{\ensuremath{{\mu}}}}
\newcommand{\nn}{{\ensuremath{{\nu}}}}
\newcommand{\kk}{{\ensuremath{{\lambda}}}}
\newcommand{\rr}{{\ensuremath{{\rho}}}}

\newcommand{\fud}[2]{{}^{#1}{}_{#2}\,}
\newcommand{\fdu}[2]{{}_{#1}{}^{#2}\,}

\newcommand{\tr}{{\mathrm{tr}}}

\newcommand{\action}[2]{{\left\langle\vphantom{#2}#1\,\right|\left.\vphantom{#1}#2\right\rangle}}
\newcommand{\scalar}[2]{{\left\langle\vphantom{#2}#1\right.;\left.\vphantom{#1}#2\right\rangle}}

\newcommand{\besubeqs}{\begin{subequations}}
\newcommand{\esubeqs}{\end{subequations}}

\usepackage{tensor}
\usepackage{todonotes}

\renewcommand{\bar}[1]{\overline{#1}}

\makeatletter
\newcommand{\subalign}[1]{%
  \vcenter{%
    \Let@ \restore@math@cr \default@tag
    \baselineskip\fontdimen10 \scriptfont\tw@
    \advance\baselineskip\fontdimen12 \scriptfont\tw@
    \lineskip\thr@@\fontdimen8 \scriptfont\thr@@
    \lineskiplimit\lineskip
    \ialign{\hfil$\m@th\scriptstyle##$&$\m@th\scriptstyle{}##$\crcr
      #1\crcr
    }%
  }
}
\makeatother

\begin{document}
\pagenumbering{gobble}
\hfill
\vskip 0.01\textheight
\begin{center}
{\Large\bfseries 
Actions for Self-dual Higher Spin Gravities}

\vskip 0.03\textheight
\renewcommand{\thefootnote}{\fnsymbol{footnote}}
Kirill \textsc{Krasnov}${}^{a}$ \&  Evgeny \textsc{Skvortsov}\footnote{Research Associate of the Fund for Scientific Research -- FNRS, Belgium}${}^{b,c}$ \& Tung \textsc{Tran}${}^{b}$
\renewcommand{\thefootnote}{\arabic{footnote}}
\vskip 0.03\textheight

{\em ${}^{a}$School of Mathematical Sciences, \\University of Nottingham, NG7 2RD, UK}\\
\vspace*{5pt}
{\em ${}^{b}$ Service de Physique de l'Univers, Champs et Gravitation, \\ Universit\'e de Mons, 20 place du Parc, 7000 Mons, 
Belgium}\\
\vspace*{5pt}
{\em ${}^{c}$ Lebedev Institute of Physics, \\
Leninsky ave. 53, 119991 Moscow, Russia}\\

\vskip 0.02\textheight

\begin{abstract}
Higher Spin Gravities are scarce, but covariant actions for them are even scarcer. We construct covariant actions for contractions of Chiral Higher Spin Gravity that represent higher spin extensions of self-dual Yang-Mills and self-dual Gravity theories. The actions give examples of complete higher spin theories both in flat and (anti)-de Sitter spaces that feature gauge and gravitational interactions. The actions are based on a new description of higher spin fields, whose origin can be traced to early works on twistor theory. The new description simplifies the structure of interactions. In particular, we find a covariant form of the minimal gravitational interaction for higher spin fields both in flat and anti-de Sitter space, which resolves some of the puzzles in the literature.
\end{abstract}

\end{center}
\newpage
\tableofcontents
\newpage
\section*{Introduction}
\pagenumbering{arabic}
\setcounter{page}{1}
Higher Spin Gravities (HiSGRA) --- theories that extend gravity with massless higher spin states --- should be good probes of the Quantum Gravity problem because of several reasons: as string theory and $AdS/CFT$ suggest, higher spin states, whether massless or massive, should be important for UV completion of gravity, while the masslessness can simulate some properties of the UV regime already at the classical level. A posteriori the latter might explain numerous difficulties encountered by HiSGRA: it is not easy to construct even a toy quantum gravity model. In any event, there is only a handful of HiSGRA, all of which exist thanks to certain subtleties in the no-go theorems.

Purely massless, partially-massless and conformal HiSGRA in three dimensions \cite{Blencowe:1988gj,Bergshoeff:1989ns,Campoleoni:2010zq,Henneaux:2010xg,Pope:1989vj,Fradkin:1989xt,Grigoriev:2019xmp} can always be formulated as Chern-Simons theories with an additional data \cite{Grigoriev:2020lzu} and they do not have propagating degrees of freedom. Conformal HiSGRA in $4d$ \cite{Segal:2002gd,Tseytlin:2002gz,Bekaert:2010ky} is a nice example of a pertubatively local HiSGRA, which extends conformal (Weyl) gravity. There is just one class of HiSGRA, Chiral HiSGRA, \cite{Metsaev:1991mt,Metsaev:1991nb,Ponomarev:2016lrm,Skvortsov:2018jea,Skvortsov:2020wtf} that features propagating massless higher spin fields and admits an action. However, the action has been so far known only in the light-cone gauge. Our paper is a step towards a covariant formulation of Chiral HiSGRA. It also aims at resolving some old puzzles and confusions in the literature regarding (non)existence of minimal gravitation interactions of higher spin fields and the (ir)relevance of the cosmological constant. 

Chiral HiSGRA can be interesting for a number of reasons: (i) it has a simple action; (ii) it is shown to be UV-finite at one-loop \cite{Skvortsov:2018jea,Skvortsov:2020wtf,Skvortsov:2020gpn} and it is expected to be one-loop exact; (iii) it captures a subset of correlation functions of Chern-Simons matter theories \cite{Skvortsov:2018uru} (all of them at the $3$-point level); (iv) its one-loop amplitudes in flat space contain all helicity plus one-loop amplitudes of perturbative QCD \cite{Skvortsov:2020gpn}; (v) Chiral HiSGRA is very close to self-dual Yang-Mills (SDYM) and self-dual Gravity (SDGRA) theories \cite{Ponomarev:2017nrr}, which are of great importance on their own. This link to SDYM and SDGRA is what is exploited in this paper. 

The main goal of the present paper is to construct covariant actions for HiSGRA with propagating massless fields. The theories we construct are close relatives of Chiral HiSGRA and correspond to its simple contractions \cite{Ponomarev:2017nrr}. These contractions are directly related to SDYM and SDGRA and can be thought of as their extensions to higher spins. For that reason, we refer to them as HS-SDYM and HS-SDGRA. Importantly, the actions of HS-SDYM and HS-SDGRA are complete gauge invariant actions, and feature interaction vertices up to the quintic order in the fields. This should be contrasted to previous results in the literature that are mostly devoted to construction of specific low order interactions. 

The actions we construct are based on a new description of free massless higher spin fields, the roots of which can be traced back to twistor theory. One of the advantages of the new approach is that it admits gauge and gravitational interactions of higher spin fields, which have long been known to exist in the light-cone or spinor-helicity approach \cite{Bengtsson:1983pd,Bengtsson:1986kh,Benincasa:2007xk,Conde:2016vxs}, but whose formulation in terms of widely adopted Fronsdal fields seems impossible \cite{Berends:1984rq,Boulanger:2006gr,Zinoviev:2008ck,Manvelyan:2010je}. The new approach allows one to uplift the (previously known only in light-cone gauge) gauge and gravitational interactions of higher spin fields to Lorentz invariant off-shell expressions both in flat and $AdS_4$ spaces. The vertices have a smooth flat limit, i.e. the new approach does not call for the cosmological constant. Also, the new description of higher spin fields is well-adapted to a much richer class of backgrounds. In particular, what can be called self-dual backgrounds are treated on an equal footing with the maximally symmetric backgrounds such as A(dS).

The self-dual theories that this paper is about (SDYM, SDGRA and we believe HS-SDYM, HS-SDGRA and Chiral HiSGRA as well) contain important information themselves: (i) self-dual backgrounds are much richer than just flat or (anti)-de Sitter spaces, over which the usual perturbation theory is formulated; (ii) certain features of complete theories can be understood already from their self-dual truncations, e.g. all helicity plus one-loop amplitudes of perturbative QCD coincide with those of SDYM (similarly for SDGRA vs. Gravity); the two-loop divergence of pure Gravity can be traced back to a non-vanishing amplitude, $A^{++++}$, of SDGRA, see \cite{Bern:2017puu}; (iii) the twistor space methods are well-suited for self-dual theories; (iv) full Yang-Mills and Gravity theories can be understood as perturbations of their self-dual sectors, see e.g. \cite{Krasnov:2016emc}, which leads both to new insights into their structure and to new computational techniques, see e.g. \cite{Chicherin:2014uca,Adamo:2020yzi}; (v) the self-dual theories have much better UV-properties, see \cite{Krasnov:2016emc}. 

We expect that most of the features listed above can be extended to Chiral HiSGRA, except for the fact that the complete HiSGRA, of which the chiral one is supposed to be a contraction, requires new ideas that go beyond the standard field theory approach due to non-locality \cite{Bekaert:2010hp,Fotopoulos:2010ay,Boulanger:2015ova,Ponomarev:2017nrr,Roiban:2017iqg,Bekaert:2015tva,Sleight:2017pcz,Ponomarev:2017qab}. In particular, we expect that there should exist a covariant action for Chiral HiSGRA, both in flat and anti-de Sitter spaces, that has a smooth flat limit. Our results also suggest that the twistor space techniques, of which in this paper we see only the final results in spacetime, should provide the most natural way to construct interactions of higher spin fields.

The paper organized as follows. In section \ref{sec:free} we discuss various formulations of free higher spin fields. In particular, we introduce a new description of higher spins and discuss its precursors. In section \ref{sec:HSYM} we, after a brief overview of SDYM, construct its higher spin extension. In section \ref{sec:HSGR} we construct a higher spin extension of SDGRA, both in flat and anti-de Sitter spaces. The relation between the results of this paper and previously available results on higher spin interactions are discussed in section \ref{sec:discussion}.

\section{Free Higher Spin Fields}
\label{sec:free}

The spin one field (gauge field) is usually described by an object with a single spacetime index. Similarly, the spin two field (graviton) is described by the metric perturbation, which is an object with two spacetime indices. It is therefore not surprising that the standard Fronsdal description of higher spin fields, see below, is based on a generalisation of this and uses objects with many spacetime indices. However, it has been known for a long time that there exists an alternative description of (free) higher spin fields. We will first review the standard Fronsdal approach, and then motivate and develop the alternative description. 
\subsection{Fronsdal approach to higher spin fields.}

It is customary in the higher spin world to describe a massless spin-$s$ field by Fronsdal tensor \cite{Fronsdal:1978rb} $\Phi_{\mm_1...\mm_s}(x)$ that is totally-symmetric, yet obeys a strange double-trace constraint $\Phi\fud{\nn\kk}{\nn\kk\mm_5...\mm_s}\equiv0$. The Fronsdal field is a gauge field
\begin{align}
    \delta \Phi_{\mm_1...\mm_s}&= \nabla_{\mm_1} \xi_{\mm_2...\mm_s} +\text{permutations}\,,
\end{align}
where the gauge parameter is traceless $\xi\fud{\nn}{\nn\mm_3...\mm_{s-1}}\equiv0$. Here $
\nabla_\mm$ is the Levi-Civita connection with respect to the same background metric that was implicitly involved in the trace constraints. Fronsdal tensor is a generalization of the gauge potential, $A_\mm$, and of the metric tensor, $g_{\mm\nn}$ to any spin. The algebraic constraints take effect starting from $s=3,4$ and were invisible for low spin theories, including gravity. The trace constraints involve the metric tensor, which becomes dynamical in a higher spin theory and also transforms under higher spin symmetries. This leads to various subtleties, in general. 

Fronsdal fields prefer constant curvature backgrounds. With $\Lambda$ denoting the cosmological constant, the Fronsdal equations \cite{Fronsdal:1978rb,Fronsdal:1978vb} are\footnote{Indices enclosed in the round brackets are assumed to be symmetrized by adding the minimal number of terms needed, which requires $s$, $s(s-1)$ and $s(s-1)/2$ terms, respectively.}
\begin{align*}
    \square \Phi_{\mm_1...\mm_s}- \nabla_{(\mm_1}\nabla^{\nn}\Phi_{\nn\mm_2...\mm_s)}+\tfrac12\nabla_{(\mm_1}\nabla_{\mm_s} \Phi\fud{\nn}{\nn\mm_3...\mm_s)}-m^2_s\Phi_{\mm_1...\mm_s}+2\Lambda g_{(\mm_1\mm_2}\Phi\fud{\nn}{\nn\mm_3...\mm_s)}=0\,,
\end{align*}
where $m_s^2=-\Lambda((s-2)(d+s-2)-s)$. Our convention is $[\nabla_\mm,\nabla_\nn]\xi^\kk=\Lambda \delta^\kk_\mm \xi_\nn-\Lambda \delta^\kk_\nn \xi_\mm$. As long as $\nabla_\mm$ is the connection of the de Sitter, anti-de Sitter or flat space the equation is gauge invariant. However, once the Weyl tensor $C\fud{\mm}{\nn,\kk\rr}$ is nonvanishing the equation ceases to be gauge invariant for $[\nabla,\nabla]$ brings $C\fud{\mm}{\nn,\kk\rr}$, which cannot be compensated by any other terms \cite{Aragone:1979hx}. It was mentioned in \cite{Aragone:1979hx} that (anti)self-dual backgrounds can be admissible at least for positive/negative helicity fields, which was also known from the twistor literature \cite{Penrose:1965am,Hughston:1979tq,Eastwood:1981jy,Woodhouse:1985id}. However, one cannot use Fronsdal fields for self-dual backgrounds. Higher spin fields on plane wave geometry were studied in \cite{Metsaev:1997ut}.\footnote{It is possible to put higher spin fields on higher spin flat backgrounds \cite{Sharapov:2019vyd} without facing the non-locality, but this goes beyond the present discussion. }

Fronsdal equations follow from an action \cite{Fronsdal:1978rb,Fronsdal:1978vb} that bears the same name. This is an ordinary second order action 
\begin{align}
    S&= \int \sqrt{g}\, \Phi^{\mm_1...\mm_s}\left(F_{\mm_1...\mm_s}-\tfrac12 g_{(\mm_1\mm_2}F\fud{\nn}{\nn\mm_3...\mm_s)}\right)\,,
\end{align}
where $F$ is the l.h.s of the Fronsdal equation. For some other closely related descriptions we refer to \cite{Ouvry:1986dv,Skvortsov:2007kz,Campoleoni:2012th} and references therein, thereon.

\subsection{Higher spins and representations of the Lorentz group}
To motivate an alternative description of higher spin fields, we recall that irreducible finite-dimensional representations of the Lorentz group in four dimensions are characterised by two numbers $(n,k)$. The representation space $S^{(n,k)}$ is then the space of objects with a number of unprimed, and some other number of primed spinor indices\footnote{Indices $A,B,...=1,2$ and $A',B',...=1,2$ are two-component spinor indices. They can be raised and lowered with the help of $\epsilon_{AB}=-\epsilon_{BA}$, $\epsilon_{12}=1$, idem. for $\epsilon^{AB}$, according to $\xi^A=\epsilon^{AB}\xi_B$, $\xi_B=\xi^A \epsilon_{AB}$. A four-component vector $v^\mu$ corresponds to a bi-spinor $v^{AA'}$ via Van der Waerden symbols $\sigma^\mm_{AA'}$, which in flat spacetime are Pauli matrices. Given metric $g_{\mm\nn}$ we have $g_{\mm\nn}\sigma^\mm_{AA'}\sigma^\nn_{BB'}=\epsilon_{AB}\epsilon_{A'B'}$. The covariant derivative $\nabla_{AA'} \equiv \sigma^\mm_{AA'}\nabla_{\mm}$ is defined so that $\nabla_\mm \sigma^\nn_{AA'}=0$, $\nabla_\mm \epsilon^{AB}=0$. We will use notation $e^{AA'}_\mm\equiv\sigma^{AA'}_\mm$ where $e^{AA'}_\mm$ is vierbein/tetrad. } 
\be
S^{(n,k)} \ni \psi^{A_1\ldots A_n, A'_1\ldots A'_k}\,.
\ee 
This is a vector in an irreducible representation when it is completely symmetric in its unprimed and primed indices. 

Now, the gauge field is the object $A_\mu$ with a single spacetime index, which in the spinor notation becomes the object $A^{AA'}\in S^{(1,1)}$. The (tracefree part of the) metric perturbation becomes the spinor object $h^{A_1A_2, A_1' A_2'}\in S^{(2,2)}$. It is then clear that the total spin of an object in $S^{(n,k)}$ should be defined as $(n+k)/2$. 

The convention that is often to be followed below is that all indices that are either symmetric or to be symmetrized are denoted by the same letter with the number of indices in a group indicated in brackets, e.g. $A(s)\equiv A_1...A_s$. This simplifies formulas, especially in the higher spin case. In this notation, an object in $S^{(n,k)}$ becomes written as
\be
S^{(n,k)} \ni \psi^{A(n), A'(k)}\,.
\ee
The symmetrization is defined to be a projector, e.g. $\xi^A \eta^A= \tfrac12( \xi^{A_1}\eta^{A_2}+\xi^{A_2}\eta^{A_1})$. 

With this notation in hand, we can return to the question of description of higher spin fields. The spin one field is described by an object $A^{AA'}$, the standard description of the spin two field is with the object $h^{A(2), A'(2)}$. It is, however, clear that there exist other possible descriptions in both cases. In the case of spin one we have an object $\psi^{A(2)}$ that carries the same spin (one). In the case of spin two we have two alternative objects $A^{A(3), A'}$ and $C^{A(4)}$, as well as similar objects with mostly primed indices, all describing the spin two. 

Of course, there exists a definite reason why the standard description with $A^{AA'}$ and $h^{A(2), A'(2)}$ is special. Recall that in the Lorentzian signature the primed spinors are complex conjugates of the unprimed. This means that the "balanced" representations $S^{(n,n)}$ of the Lorentz group are the only ones that contain real vectors. This means that if we are to impose the condition that the description we are after has to be manifestly real, then the only alternative is to use the balanced representations, e.g. Fronsdal's description. 

There are, however, many benefits in dropping the manifest reality property. Indeed, the history of mathematics of the last two centuries teaches us how beneficial it may be to "complexify" the problem at hand. And indeed, both in the case of spin one and spin two we know that it is very beneficial to consider the other, inherently complex, representations of the Lorentz group.

Let us consider the spin one case first. The field strength is then a two-form, and this can be decomposed into its self- and anti-self dual parts. In the spinor formalism these two parts are described as objects $F^{A_1A_2}\equiv F^{A(2)}$ and $F^{A_1'A_2'}$, which are complex conjugates of each other. It is then well-known that both of the vacuum Maxwell's equations $d F=0$, $d {}^* F=0$ can be written as a single complex equation 
\be\label{eq-Maxwell}
 \partial_{BA'} F^{BA} =0\,.
 \ee
The two real Maxwell's equations arise as the real and imaginary parts of this complex equation. Here we have translated the operator of partial derivative $\partial_\mu$ into its spinor form $\partial_{AA'}$. We note that the equation (\ref{eq-Maxwell}) is conformally invariant.
 
Let us now consider the case of spin two. In this case we have the Riemann curvature tensor $R_{\mu\nu,\rho\sigma}$. It splits into its Lorentz irreducible components that are the Ricci scalar, the tracefree part of Ricci, and the Weyl curvature. The tracefree part of Ricci gets translated into a spinor object $\Phi^{A(2),A'(2)}$. In vacuum this vanishes. The Weyl curvature tensor $C_{\mu\nu,\rho\sigma}$ becomes described by two objects $C^{A_1 A_2 A_3 A_4}\equiv C^{A(4)}$ and $C^{A'(4)}$ that are complex conjugates of each other. In vacuum, we have a direct generalisation of (\ref{eq-Maxwell})
 \be\label{eq-Weyl}
 \nabla_{BA'} C^{BA_2 A_3 A_4} =0\,,
 \ee
where now the covariant derivative is used instead of the partial one. The complex conjugate of this equation is then an equation on $C^{A'(4)}$. This is again a conformally invariant (complex) equation. For more details on this standard material we refer the reader to the book \cite{Penrose:1987uia}. 
 
The pattern in (\ref{eq-Maxwell}), (\ref{eq-Weyl}) is clear and there is a direct generalisation of these equations to the higher spin case \cite{Penrose:1965am}
\be\label{eq-higher}
\nabla_{BA'} \Psi^{B A(2s-1)}=0\,,
\ee
where $s$ is the total spin. This is again a conformally invariant equation. The difference with Fronsdal's description based on "balanced" representations of the Lorentz group is that now one is instead using the maximally unbalanced representations, with the spinor indices of only one type. This description arises naturally in twistor theory, as we now review.

\subsection{Higher spins and twistors}
Twistor theory leads to a deep correspondence between various field equations (in complexified Minkowski space) and holomorphic structures on a three-dimensional complex manifold, the twistor space. Massless fields of any spin in $4d$ can be realized as functions on twistor space via Penrose transform. In this paper we prefer to stay in spacetime and only borrow the end results of twistor theory. The original references are \cite{Atiyah:1979iu,Hitchin:1980hp, Eastwood:1981jy}, collection of papers \cite{Hughston:1979tq} and review \cite{Woodhouse:1985id}. Twistor space is well-suited for self-dual backgrounds. However, the propagating fields of positive \cite{Penrose:1965am,Penrose:1968me,Woodhouse:1985id} and negative \cite{Hughston:1979tq,Woodhouse:1985id} helicity are treated differently \cite{Eastwood:1981jy,Woodhouse:1985id}, which may look awkward on the first sight. Twistor constructions can be extended from free fields to self-dual Yang-Mills \cite{Ward:1977ta}, self-dual Gravity \cite{Penrose:1976js}, self-dual conformal Gravity \cite{Berkovits:2004jj} and, eventually, the twistor space description of the complete theories can be rebuilt as perturbations over the self-dual truncations \cite{Witten:2003nn,Boels:2006ir,Adamo:2013tja}. Twistor techniques were applied to conformal HiSGRA in \cite{Adamo:2016ple,Haehnel:2016mlb}.

As we already reviewed, a massless spin-$s$ field on Minkowski or conformally-flat background (e.g. anti-de Sitter) can be described by two rank-$2s$ spin-tensors $\Psi^{A_1...A_{2s}}$, $\Psi^{A_1'...A_{2s}'}$, see e.g. \cite{Penrose:1965am}, which are complex conjugate of each other
and obey 
\begin{align}\label{hsA}
    \nabla^\fdu{B}{A'} \Psi^{BA(2s-1)}&=0\,, &\nabla\fud{A}{B'} \Psi^{B'A'(2s-1)}&=0\,,
\end{align}
where we used the already described convention that the symmetrised spinor indices are denoted by the same letter. However, this is not how massless fields emerge from twistor theory. The first equation can be left as it is and describes, say, positive helicity fields. The opposite helicity requires a different approach \cite{Hughston:1979tq,Eastwood:1981jy,Woodhouse:1985id}. It is described by a gauge potential $\Phi^{A_1...A_{2s-1},A'}$ that obeys
\begin{align}\label{hsB}
    \nabla\fud{A}{A'}\Phi^{A(2s-1),A'}&=0\,, && \delta \Phi^{A(2s-1),A'}=\nabla^{A A'}\xi^{A(2s-2)}\,.
\end{align}
The self-consistency (gauge invariance for \eqref{hsB})  of the equations above requires
\begin{align}\label{symA}
    [\nabla\fud{A}{C'},\nabla^{AC'}]\chi^{A}&=0\,.
\end{align}
Eq. \eqref{symA} must hold for any spinor $\chi^A$, which implies that the self-dual (SD) component of the Weyl tensor vanishes, $C_{ABCD}=0$. In particular, all constant curvature backgrounds are admissible. Moreover, the equation becomes well-posed on an arbitrary self-dual spacetime,\footnote{There are no self-dual spacetimes apart from the maximally symmetric ones if one requires the signature to be Lorentzian. It is then customary in twistor theory to go beyond Minkowski signature by considering complexified spacetimes, Euclidian and split signatures and/or by treating $\Psi$ as (complex) wave functions that carry positive energy. Such extensions beyond Minkowski are also necessary for the modern amplitude techniques, as this uses a three-point amplitude, which vanishes for spinning fields in Minkowski. In what follows we assume that one or another point of view is chosen and will not discuss reality/positive energy conditions explicitly. } a fact known from 70's. However, SD spacetimes are not reachable via Fronsdal equation: the gauge parameter of the Fronsdal field $\xi_{A(s-1),A'(s-1)}$ has both types of indices and $[\nabla,\nabla]$ will produce both SD and ASD Weyl tensors upon taking the gauge variation.

\subsection{The chiral connection formalism for GR}
\label{subsec:GR}

There is yet another story where the linearised description based on (\ref{hsA}), (\ref{hsB}) arises. This is the chiral "pure connection" formalism for General Relativity developed in the series of works \cite{Krasnov:2011up,Krasnov:2011pp,Krasnov:2012pd,Delfino:2012zy,Delfino:2012aj,Fine:2013qta,Delfino:2014xea,Krasnov:2014wha}. The link to the developments of this paper is that the description of self-dual GR \cite{Krasnov:2016emc} that is here generalised to higher spins is based precisely on the "connection" variables.

The link to (\ref{hsA}), (\ref{hsB}) is as follows. First, there exists a chiral "pure spin connection" formulation of GR, first exhibited in \cite{Krasnov:2011pp}, and arrived at by the procedure of "integrating out" the metric-like and the Lagrange multiplier fields from the Plebanski action \cite{Plebanski:1977zz}. As is described explicitly in \cite{Krasnov:2012pd}, on any self-dual background, the linearisation of the GR Lagrangian in this formalism takes the form $(\nabla\fdu{A}{B'}\Phi_{A(3),B'})^2$. This can be written in a first-order form in which an auxiliary field $\Psi^{A(4)}$ is introduced so that the linearised Lagrangian becomes
\be
L\sim \Psi^{A(4)} \nabla\fdu{A}{B'}\Phi_{A(3),B'} - \frac{1}{2} (\Psi^{A(4)})^2\,.
\ee
The first of these terms appears as the linearised Lagrangian of self-dual GR \cite{Krasnov:2016emc}. The second term gets added when one passes from the theory of SDGRA to the full GR, see \cite{Krasnov:2016emc}. In both cases the higher spin generalization is straightforward. 

In this "pure connection" formalism for GR, the field $\Phi_{A(3), B'}$ arises as the projection of a one-form field $\omega_{A_1A_2}\equiv \omega_{A_1A_2; BB'}\, dx^{BB'}$ onto its totally symmetric part $\omega_{A(3), A'}$ in its unprimed spinor indices. The connection field $\omega_{A_1A_2; BB'}\, dx^{BB'}$ is a one-form with values in the Lie algebra of the self-dual "half" ${\rm SO}(3,{\mathbb C})$ of the Lorentz group. The projection onto the totally symmetric part arises because the theory is invariant not only under gauge ${\rm SO}(3,{\mathbb C})$ rotations, but also under diffeomorphisms. In the "pure connection" setup under consideration, the diffeomorphisms act purely algebraically, by shifts\footnote{\label{ft:Cartan}Let us recall Cartan formula $\mathcal{L}_\eta=\{d,i_\eta\}$, which can be used to represent Lie derivative $\mathcal{L}_\eta A$ of a connection $A$ as $\mathcal{L}_\eta A=D(i_\eta A)+i_\eta F$, i.e. as a combination of a gauge transformation and of the curvature term.} 
\be\label{diffeos-spin-two}
\delta_\eta  \omega_{A_1A_2} = i_\eta H_{A_1 A_2}\,,
\ee
where we introduced the self-dual 2-forms $H^{AB}\equiv e\fud{A}{C'}\wedge e^{BC'}$, and $i_\eta$ is the operation of insertion of a vector field (interior product or inner derivative). We refer the reader to the series of papers \cite{Krasnov:2011up,Krasnov:2011pp,Krasnov:2012pd,Delfino:2012zy,Delfino:2012aj,Fine:2013qta,Delfino:2014xea,Krasnov:2014wha} for more details. In particular, already the first paper \cite{Krasnov:2011up} emphasises the importance of this realisation of the action of diffeomorphisms on the basic connection field. We now generalise this description of free spin two fields to higher spins.
\subsection{A simple action for free higher spin fields.} 
The equations \eqref{hsA} and \eqref{hsB} follow from a simple action
\begin{align}\label{actionA}
    S&=\int \sqrt{g}\, \Psi^{BA_2...A_{2s}} \nabla\fdu{B}{B'}\Phi_{A_2...A_{2s},B'}
\end{align}
This action was almost written in \cite{Hitchin:1980hp}, where it was noted that the adjoint operator of \eqref{hsB} corresponds to \eqref{hsA}.\footnote{This is a math's way of saying 'there is an action'. For the case of flat space this action was also discussed in \cite{Flores:2017yyj}. We are grateful to D.Ponomarev for bringing this paper to our attention. } As reviewed in the previous subsection, for the case of spin two, this action arises naturally in the "connection" description of Einstein 4-manifolds. In particular, the paper \cite{Krasnov:2014wha} contains a rather general discussion of issues arising in the case of spin two. 

As in the case of the "connection" description of spin two, and in the general higher spin case discussed in \cite{Hitchin:1980hp}, there is an even more geometric reformulation. Thus, one introduces a one-form connection 
\be
\omega^{A_1...A_{2s-2}}=\Phi^{A_1...A_{2s-2}B,B'} dx_{BB'}\,,
\ee
and relaxes the identification with $\Phi$. As a result, $\omega^{A_1...A_{2s-2}}$ acquires one more irreducible component that we will need to take care of, i.e. to make sure it does not lead to any propagating degrees of freedom. The decomposition into irreducible spin-tensors reads
\begin{align}
    \omega^{A(2s-2)}\equiv e_{BB'}\Phi^{A(2s-2)B,B'}+e\fud{A}{B'}\Theta^{A(2s-3),B'}\,,
\end{align}
where $e^{AA'}\equiv e^{AA'}_\mm \, dx^\mm$ is the vierbein one-form.
For $s=1,2$ we recover the Maxwell gauge potential $\omega$ and the self-dual part of the spin-connection that plays the main role in the description of \cite{Krasnov:2016emc}. The gauge transformations take the form\footnote{We note that the same and more general type of variables, one-forms $\omega^{A(n),A'(m)}$, were introduced in \cite{Vasiliev:1986td} as well as zero-forms $C^{A(n),A'(m)}$ of general type. This formulation contains all possible types of fields that can be relevant for description of higher spin fields at the level of equations of motion. }
\begin{align}\label{lin-gauge}
    \delta \omega^{A(2s-2)}&= \nabla \xi^{A(2s-2)} +e\fud{A}{C'} \eta^{A(2s-3),C'}\,,
\end{align}
where the first term represents the usual "gauge" invariance, while the second term projects out the unwanted components of $\omega^{A(2s-2)}$ and generalises (\ref{diffeos-spin-two}). It is then clear that the following action\footnote{Curiously enough, this action can be obtained as a particular case of the presymplectic AKSZ action of \cite{Sharapov:2021drr}.}
\begin{align}\label{niceaction}
    S=\int \Psi^{A_1\ldots A_{2s}}\wedge H_{A_1A_2}\wedge \nabla \omega_{A_3\ldots A_{2s}}\equiv \int \Psi^{A(2s)}\wedge H_{AA}\wedge \nabla \omega_{A(2s-2)}\,.
\end{align}
is gauge invariant and, hence, equivalent to \eqref{actionA}. Indeed, for SD backgrounds we have $\nabla^2\xi_{A(2s-2)}\equiv \Lambda H\fdu{A}{B}\xi_{BA(2s-3)}$, where $\Lambda$ is some constant, and  the variation vanishes thanks to $H_{AA}\wedge H\fdu{A}{B}\equiv0$. The algebraic symmetry, which is the second term in (\ref{lin-gauge}) is valid thanks to $H_{AA}\wedge e_{AB'}\equiv0$, of which $H_{AA}\wedge H\fdu{A}{B}\equiv0$ is a simple consequence. 

Action \eqref{niceaction} is, perhaps, the simplest action for massless higher spin fields on Minkowski and (anti)-de Sitter backgrounds. It also works for self-dual spaces. We note that, as in any "chiral" (complexified) formalism, the positive and negative helicities are treated differently. The action is also valid for fermionic (anti-commuting) fields, which have half-integer $s$ and odd number of $A$ indices, but we only discus the case of bosonic fields. It is worth stressing that actions \eqref{actionA} and \eqref{niceaction} are valid actions for $s\geq1$, i.e. they do not cover the matter fields with $s=0,\tfrac12$. 

Lastly, in parallel with the discussion of GR Lagrangian in section \ref{subsec:GR} there is a simple second order Lagrangian for higher spin fields in terms of $\Phi$
\begin{align}
    \mathcal{L}&= (\nabla\fdu{A}{B'}\Phi_{A(2s-1),B'})^2 \sim \Psi^{A(2s)}(\nabla\fdu{A}{B'}\Phi_{A(2s-1),B'})-\tfrac12 (\Psi_{A(2s)})^2\,,
\end{align}
which can be written in the first order form with the help of $\Psi^{A(2s)}$. Action \eqref{actionA} is obtained by dropping the last term. The second order equations resulting from the action are gauge invariant on SD backgrounds and describe both helicities. This is the second (new) description of free higher spin fields. In this paper we employ the first one, which is based on \eqref{niceaction}, to construct examples of interacting higher spin theories.

\subsection{Generating functions} 
For future convenience, let us rewrite the action in an index-free form with the help of generating functions. Firstly, we pack all connections as
\begin{align}
    \omega\equiv \omega(y|x)&= \sum_k \frac{1}{k!}\omega_{A_1...A_k }\, y^{A_1}...\, y^{A_k} \,,
\end{align} 
where $y^A$ is a commuting auxiliary variable and, likewise, for $\Psi=\Psi(y|x)$. 
Formally, we deal with sections of $S(S_+) \otimes \Omega(M)$, which is a product of the symmetric tensor algebra of the spinor bundle and the algebra of differential forms $\Omega(M)$ or, equivalently, we can consider the algebra $C[y]\otimes\Omega(M)$ of polynomials (or formal series) in $y^A$ times $\Omega(M)$. For two elements $X$, $Y$ such that the total differential form degree equals the spacetime dimension $d$ we define
\begin{align}\label{scalarp}
    \action{X}{Y}&= \int \scalar{X}{Y}\,, &
    \scalar{X}{Y}&= \sum_n \frac{1}{n!} X^{A(n)} \wedge Y_{A(n)}\,.
\end{align}
With $\Psi=\Psi(y|x)$ being the generating function of $\Psi^{A(2s)}$ the free actions sum up to
\begin{align}
    S&= \action{\Psi}{\tfrac12 H_{AA}y^Ay^A \wedge \nabla \omega}\,,
\end{align}
where $\nabla$ acts component-wise. The gauge transformations can be represented as
\begin{align}
    \delta \omega&= \nabla \xi+ \overrightarrow{i_\eta}(\tfrac12 H_{AA}y^Ay^A)\,,
\end{align}
where instead of $\eta^{A(2s-1),A'}$ we define generalized vector fields $\eta^\mm(y)$ as sections of $C[y]\otimes TM$. For any differential form $X=X(y;x;dx)$ the (new) inner derivative is defined as
\begin{align}\label{innerR}
    \overrightarrow{i_\eta} X&= \eta^\mm (y) \frac{\pl}{\pl (dx^\mm)} X\,,
\end{align}
which is, thereby, inert to the spinor indices hidden under $y^A$. Therefore, we can define $\eta^{A(2s-1),A'}$ as follows
\begin{align}\label{tworep}
    {e_\mm}\fud{A}{C'}\eta^{A(2s-1),C'}= \eta^{A(2s-2)|\nn} H^{AA}_{\nn\mm}\,.
\end{align}
Note that $\eta^\mm(y)$ contains another irreducible component not present in (\ref{lin-gauge}), but this gets projected out due to symmetrization.

\section{Higher Spin SDYM}
\label{sec:HSYM}
The first class of higher spin theories that we construct extends the SDYM action. We start by reviewing the latter. 
\subsection{Self-Dual Yang-Mills}
Our starting point is the covariant Chalmers-Siegel action \cite{Chalmers:1996rq} for SDYM, which we write as
\begin{align}
    S_{SDYM}&=\tr \int B \wedge F\,,
\end{align}
where $F=dA+\tfrac12 [A,A]$ is the Yang-Mills field-strength of connection $A$ for some Lie algebra $\mathfrak{g}$, and $B$ is a self-dual two-form valued in $\mathfrak{g}$. Without loss of generality we assume that $\mathfrak{g}$ is realized as a matrix algebra and $\tr$ is the (invariant) trace. With the help of the basis self-dual two-forms $H^{AA}$, $H^{AB}\equiv e\fud{A}{C'}\wedge e^{BC'}$ we can represent $B$ as $B=\Psi^{AA} H_{AA}$ for some zero-form $\Psi^{AA}$, while $F$ can be decomposed as
\begin{align}
    F&= H_{AA} F^{AA} +H_{A'A'} F^{A'A'}\,, && F_{AA'|BB'}=\pl_{A'A} A_{BB'}-\pl_{B'B} A_{AA'} +[A_{AA'},A_{BB'}] \,,
\end{align}
where $F_{AB}\equiv F_{AA'|BB'}\epsilon^{A'B'}$, $F_{A'B'}\equiv F_{AA'|BB'}\epsilon^{AB}$. As a result, the action can be written as
\begin{align}\label{SDYM}
    S_{SDYM}&=\tfrac12\tr \int \Psi^{AA} \wedge H_{AA} \wedge F\,.
    \end{align}
 Using the generating functional we can rewrite this as\footnote{The definition of the pairing $\action{X}{Y}= \tr \int \scalar{X}{Y}$ implicitly incorporates the trace.}
  \begin{align}  
  S_{SDYM}&= \action{\tfrac12\Psi_{AA}y^Ay^A}{\tfrac12 H_{AA}y^Ay^A \wedge F}
\end{align}
and its free approximation agrees with \eqref{niceaction} for $s=1$
\begin{align}
    S_{free}&=\tfrac12\tr \int \Psi^{BB} \wedge H_{BB} \wedge dA    \,.
\end{align}

\subsection{Higher spin extension} 
Let us consider flat, (anti)-de Sitter or even any self-dual gravitational background. The spacetime is described by covariant derivative $\nabla$ and vierbein $e^{AA'}$. The latter defines the basis $H^{AA}$ of self-dual two-forms. It is easy to construct a higher spin extension of the SDYM action. We take one-form $\omega^{A(2s-2)}$ for every $s$ and pack them into a generating function $\omega(y|x)$. In order to switch on Yang-Mills gauging we assume $\omega(y|x)$ to take values in a Lie algebra $\mathfrak{g}$, i.e. the components are one-forms valued in some Lie algebra. A bit more formally connection $\omega$ takes values in $C[y]\otimes \Omega(M)\otimes \mathfrak{g}$. As before, we assume that $\mathfrak{g}$ is realized in $\mathrm{Mat}_N$ for some $N$ and, hence, $\omega(y|x)\equiv \omega(y|x)\fdu{i}{j}$ are $N\times N$ matrices and we write $\omega\wedge \omega$ instead of $\tfrac12 [\omega,\omega]$.

The field-strength is defined in the standard way, $F=d\omega+\omega\wedge \omega$ (or with any self-dual $\nabla$ instead of $d$), and contains the commutator of the components followed by the symmetrization over all spinor indices, which is achieved automatically thanks to $y^A$:\footnote{Here and below all fields and gauge parameters are assumed to have even number of indices, i.e. are bosons. A generalisation to fermions is straightforward. }
\begin{align}
    \omega \wedge \omega &= \sum_{n,m=0} \frac{1}{2\,n!m!}\,[\omega_{A(n)}, \omega_{A(m)}]\, y^{A_1}...\,y^{A_{n+m}}\,.
\end{align}

The higher spin action we consider is the direct generalisation of (\ref{SDYM}) to higher spin, which is achieved by allowing all objects to take values in polynomials in the $y$ variables. The action can be written in several equivalent forms. We have (now, $\action{\bullet}{\bullet}$ is improved with $\tr$):
\begin{align}\label{HSSDDYM}
    S&=\action {\Psi}{\tfrac12 H_{AA}y^Ay^A \wedge F }=\sum_{s=1}\tfrac{1}{(2s)!}\tr\int \Psi^{A(2s)} \wedge H_{AA}\wedge F_{A(2s-2)}\,.
\end{align}
This action is invariant under the usual Yang-Mills transformations:
\begin{align}
    \delta \omega&= \nabla\xi+[\omega,\xi]& 
    \delta \Psi&= [\Psi,\overleftarrow{\xi}]=\sum_{i,n}\tfrac{1}{n!}[\Psi\fdu{A(n)}{B(i)},\xi_{B(i)}]\,,
\end{align}
where $\xi$ is a generating function $\xi(y |x)$ and the commutator in the last expression is the usual matrix commutator. It is important that the commutator does not touch the spacetime components, which are eventually associated with expansion in $y^A$. 

The higher spin extension of (\ref{SDYM}) introduces a new type of gauge invariance, not present in the usual SDYM. Indeed, the action is also invariant under 
\begin{align}\label{symmetry-shifts}
    \delta \omega^{A(k)}&= e\fud{A}{C'} \eta^{A(k-1),C'}\,,
\end{align}
which is again a simple consequence of $H_{AA}e_{AB'}\equiv0$. 

Lastly, we would like to stress that the Yang-Mills type interactions of (higher spin) fields are non-abelian ones, i.e. they do deform the gauge algebra and, hence, the consistency of the action is far from being trivial.

\subsection{Gauge-fixing and amplitudes}
\label{subsec:ampl}
The action (\ref{HSSDDYM}) contains the kinetic term of the schematic type $\Psi \partial \omega$, as well as the cubic interaction terms $\Psi \omega\omega$. The interaction can be usefully characterised in terms of amplitudes. 

For this purpose, we first discuss how the gauge symmetry present in the kinetic term can be gauge fixed. We then present the polarisation states describing higher spin fields of positive and negative helicities. Such states can then be inserted into the cubic vertex to obtain the 3-point amplitudes that this vertex describes. 

As is appropriate in the amplitude context, we consider a flat background. The gauge-fixing of the symmetry (\ref{symmetry-shifts}) is done by simply requiring the (linearised) connection field $\omega_{A(2s-2);BB'}$ to take values in the irreducible Lorentz representation $\Phi_{A(2s-1),A'}$. Given that the gauge transformation (\ref{symmetry-shifts}) does not contain derivatives of the gauge transformation parameter, the algebraic gauge-fixing employed is appropriate. The arising kinetic term is then
\be\label{kinetic}
\Psi^{A(2s)} \partial_{A A'} \Phi_{A(2s-1),}{}^{A'}\,.
\ee
This kinetic term is still invariant under the $\delta\Phi_{A(2s-1)A'} = \partial_{AA'}\xi_{A(2s-2)}$ gauge-transformations, because $\partial_{AA'}\partial_{B}{}^{A'}\sim \square \epsilon_{AB}$, which then vanishes after the symmetrization. This kinetic terms also exhibit a mismatch in the fields. The dimension of the representation described by an object $\psi_{A(k)}$ is $k+1$, and so there are $4s$ components in the field  $\Phi_{A(2s-1),A'}$, while there are only $2s+1$ components in the field $\Psi^{A(2s)}$. The mismatch $4s-(2s+1)=2s-1$ is precisely the number of the components in the gauge transformation parameter $\xi_{A(2s-2)}$. 

This gauge symmetry can be fixed by adding the generalised Lorentz gauge condition $\partial^{BA'} \Phi_{BA(2s-2), A'}=0$. This gauge condition can be added to the action with a Lagrange multiplier field, as a term
\be\label{gauge-fixing}
\chi^{A(2s-2)} \partial^{BA'} \Phi_{BA(2s-2), A'}\,.
\ee
It can then be noted that the terms (\ref{kinetic}) and (\ref{gauge-fixing}) can be put together as
\be
\tilde{\Psi}^{A(2s-1); B} \partial_{B A'} \omega_{A(2s-1)}{}^{A'}\,,
\ee
where we have defined a new field
\be
\tilde{\Psi}^{A(2s-1); B} = \Psi^{A(2s-1) B}+  \chi^{A(2s-2)} \epsilon^{AB}\,.
\ee
The number of components in the new field $\tilde{\Psi}^{A(2s-1); B}$ matches the number of components in $\Phi_{A(2s-1),}{}^{A'}$ and the kinetic term is completely gauge-fixed. The operator appearing in the kinetic term is then a version of the chiral Dirac operator. Its inverse is then the propagator that arises in this theory. 

One of the two helicities described by (\ref{HSSDDYM}) resides in the connection field, while the other helicity resides in $\Psi^{A(2s)}$. Let us agree that it is the negative helicity that is described by the connection. The corresponding polarisation tensor is then
\be
\epsilon^-_{A(2s-1),A'}(k) = M^{s-1} \frac{q_{A_1} \ldots q_{A_{2s-1}} k_{A'}}{ (qk)^{(2s-1)}}\, .
\ee
Here we have a null momentum $k_{AA'}=k_{A} k_{A'}$, $q_A$ is an auxiliary spinor, and $(qk):= q^A k_A$ is the spinor contraction. We have also included a dimensionful parameter $M$ to an appropriate power in front, so that the polarisation spinor is dimensionless for all $s$.
The case of $s=1$ gives the familiar polarisation spinor of YM theory. The case $s=2$ reproduces the states considered in \cite{Delfino:2012aj}. We note that the polarisation spinors introduced satisfy the (momentum space version of the) linearised field equation
\be
k^{AA'} \epsilon^-_{A(2s-1),A'}(k) =0\,.
\ee

The polarisation spinor describing the opposite, positive helicity is an object with only unprimed spinor indices and is given by
\be
\epsilon^+_{A(2s)} = M^{1-s} k_{A_1} \ldots k_{A_{2s}} \,.
\ee
The mass parameter in front is so that the mass dimension of this spinor is always one. In particular, for $s=1$ there is no extra mass parameter that is needed. This polarisation spinor satisfies the equation
\be
k^{BA'} \epsilon^-_{B A(2s-1)}=0\,.
\ee

We can finally characterise the cubic vertex appearing in (\ref{HSSDDYM}). It is clear that it gives a pairing of two plus and one minus helicity states. The self-dual projection of the two-form $\omega\omega$ evaluated on two positive helicity states of momenta $1,2$ and spins $s_1, s_2$ is given by
\be
\omega\omega \sim M^{s_1+s_2-2} \frac{[12]}{ (q1)^{(2s_1-1)} (q2)^{(2s_2-1)}  } q_{A_1} \ldots q_{A_{2s_1-1}} q_{B_1} \ldots q_{B_{2s_2-1}}\,,
\ee
where $[12]:=k_1^{A'} k_{2 A'}$ is the contraction of the corresponding primed spinors. This quantity can now be paired with the negative polarisation spinor of momentum $k_3: k_1+k_2+k_3=0$. The total spin of the state must be such that the unprimed indices match, and so $2s_3= 2s_1 + 2s_2 -2$. This means that $s_1+s_2-2=s_3-1$, which means that the dimensionful prefactor cancels out from the amplitude. This gives the following amplitude
\be
{\cal A}^{--+} = [12] \frac{ (q3)^{2s_1+2s_2-2}}{ (q1)^{(2s_1-1)} (q2)^{(2s_2-1)}  } \delta(s_3-(s_1+s_2-1))\,.
\ee
This can be further transformed using the momentum conservation to eliminate the $q$-dependence. We have $(q3)/(q1)=-[12]/[32], (q3)/(q2)=-[21]/[31]$, and so 
\be
{\cal A}^{--+} = \frac{ [12]^{2s_1+2s_2-1}}{ [23]^{(2s_1-1)} [13]^{(2s_2-1)}  } \delta(s_3-(s_1+s_2-1))\,,
\ee
which is the higher spin generalisation of the standard YM result.
\subsection{Light-cone gauge} 
Yet another way to see how the interactions between physical degrees of freedom look like is to impose the light-cone gauge:\footnote{We use $A=+,-$ instead of $A=1,2$.}
\begin{align}
    \Phi^{A(2s-2)+,+'}&=0\,,
\end{align}
which leaves us with a 'scalar' $\Phi_{+s}(x)=(\pl^{++'})^{-1}\Phi^{-...-,+'}$. The rest of the $\Phi$-components either vanish or is an auxiliary field $\Phi^{-...-,-'}=\pl^{+-'}\Phi_{+s}$. The second field $\Psi^{A(2s)}$ contains one physical component, which we denote $\Phi_{-s}=(\pl^{++'})\Psi_{-...-}$, the rest being auxiliary fields. Action \eqref{HSSDDYM}, when expressed in terms of the two physical fields, reduces to
\begin{align}\label{CSSDYM}
    \sum_s\int \tr(\Phi_{-s} \square \Phi_{+s})+g \sum_{s_1,s_2}a_{s_1,s_2}\epsilon_{A'B'} \tr\left(\Phi_{-(s_1+s_2-1)} [\pl^{+A'}\Phi_{+s_1}, \pl^{+B'}\Phi_{+s_2}] \right)\,,
\end{align}
where we recall that $\Phi_{\pm s}$ take values in a Lie algebra. Coefficients $a_{s_1,s_2}$ are not important at the moment (they have specific values as coming from the covariant action).\footnote{Here, and also in the higher-spin extension of SDGRA below, the coefficients $a_{s_1,s_2}$ are fixed, but some other choices (e.g. truncations) might exist. } The $s=1$ subsector contains the standard (chiral half of ) Yang-Mills interaction and has the well-known Chalmers-Siegel form \cite{Chalmers:1996rq}. In this regard let us note that $\Phi_{\pm s}$ are related via a simple rescaling $\phi_{\pm s}=(\pl^{+})^{\pm1}\Phi_{\pm s}$ to fields $\phi_{\pm s}$ that transform canonically under the Lorentz transformations, see e.g. \cite{Bengtsson:1986kh,Metsaev:1991mt,Metsaev:1991nb}. In particular, the interaction has only one transverse derivative, while the second derivative is $\pl^+\equiv\pl^{++}$. Therefore, in the standard picture \eqref{CSSDYM} features one-derivative interaction of Yang-Mills type. Eq. \eqref{CSSDYM} is a contraction of the Chiral HiSGRA that was considered in \cite{Ponomarev:2017nrr}.\footnote{The dictionary between \eqref{CSSDYM} and the light-cone formulas in momentum space of \cite{Ponomarev:2017nrr} is that structure $\epsilon_{A'B'} \bullet \pl^{+A'}\bullet \pl^{+B'}\bullet$ will correspond to $p^+_2 p_3-p_3^+ p_2\equiv \mathbb{P}_{23}$ in momentum space, where $p^a=(p^+,p^-, p, \bar{p})$. } In the language of \cite{Ponomarev:2017nrr} it corresponds to the commutative limit of the (kinematic) algebra. Historically, it was observed in \cite{Devchand:1996gv} that the self-dual truncation of super Yang-Mills theory does not have an upper bound for $\mathcal{N}$, which also leads to higher spin fields for $\mathcal{N}>6$.

\section{Higher Spin SDGRA}
\label{sec:HSGR}

Our other new higher spin theory is the higher spin extension of the SDGRA, which we now review. There are two formulations, one relevant in the case of a non-zero scalar curvature, the other one describing "flat" gravitational instantons. The first has been described in \cite{Krasnov:2016emc}. The second is new. Here we motivate it by a "contraction" procedure from the non-zero scalar curvature one, but it can be also described intrinsically, which will be the subject of a separate publication \cite{KrSk}. 

\subsection{SDGRA: non-zero scalar curvature}

The formulation of SDGRA with non-zero scalar curvature given in \cite{Krasnov:2016emc} is the "pure connection" one, as it uses the basic fields already encountered in section \ref{sec:free}. The dynamical fields are a zero-form $\Psi^{ABCD}$ and a connection one-form $\omega^{AB}$. The action reads
\begin{align}\label{SDGRAKirill}
    S&= \tfrac12 \int \Psi^{ABCD} \wedge F_{AB} \wedge F_{CD}\,, & F^{AB}&= d\omega^{AB} +\omega\fud{A}{C} \wedge\omega^{CB}\,.
\end{align}
It is invariant under 
\besubeqs\label{SDGRAgauge}
\begin{align}
    \delta \omega^{AB}&= d\xi^{AB} +\omega\fud{A}{C} \xi^{CB}+\omega\fud{B}{C} \xi^{CA}+\mathcal{L}_\eta \omega^{AB}\,, \\
    \delta \Psi^{ABCD}&= \Psi^{M(ABC}\xi\fdu{M}{D)} +\mathcal{L}_\eta \Psi^{ABCD}\,.
\end{align}
\esubeqs
Let us choose the constant curvature or an instanton background, i.e. we choose $\omega_0$ such that 
\begin{align}
    d\omega^{AB} +\omega\fud{A}{C} \wedge\omega^{CB}&= e\fud{A}{C'} \wedge e^{BC'}\equiv H^{AB}\,,
\end{align}
the cosmological constant being implicit on the right-hand side (we set it to $1$). The background value of $\Psi^{ABCD}$ is assumed zero. The action decomposes as
\begin{align}
\begin{aligned}
    S= \int &\Psi^{ABCD}\wedge \left (H_{AB} \wedge \nabla\omega_{CD} \right) +\\
    +&\Psi^{ABCD}\wedge \left( \tfrac12  \nabla\omega_{AB} \wedge \nabla \omega_{CD} + H_{AB}\, \omega_{CM}\wedge \omega\fdu{D}{M}\right)+\\
    +&\Psi^{ABCD}\wedge \left( \nabla\omega_{AB} \wedge \omega_{CM}\wedge \omega\fdu{D}{M} \right) \,.
\end{aligned}
\end{align}
There is a canonical free term, cf. \eqref{niceaction}. The cubic term features the $d\omega d\omega$-part and an $\omega\omega$ term due to curvature. There is also a quartic term, but the quintic one vanishes. This is the structure we would like to generalize to higher spin fields. 
\subsection{SDGRA: zero scalar curvature}
We motivate the zero scalar curvature version of (\ref{SDGRAKirill}) by referring to a "contraction" procedure that is known to give another formalism for flat space SDGRA. 

In \cite{Siegel:1992wd} the author gave a covariant formulation of flat space SDGRA that works by producing, as one of the arising Euler-Lagrange equations, the  condition that one of the chiral "halves" of the spin connection is zero. This then implies that the metric is a flat gravitational instanton. The paper \cite{AbouZeid:2005dg} notes, see formula (30) of this paper, that Siegel's action is equivalent to the action one obtains by setting to zero the $\omega\omega$ term in the chiral Plebanski action. Inspired by this, we apply the same "contraction" procedure to the action (\ref{SDGRAKirill}), to obtain an action to describe flat space SDGRA. 

The effect of the contraction procedure is to remove the $\omega\omega$ terms from the curvatures in (\ref{SDGRAKirill}). This results in the following action 
\begin{align}
    S_{\text{SDGRA, flat}}&=\tfrac12  \int \Psi^{ABCD} \wedge d\omega_{AB} \wedge d\omega_{CD} \,,
\end{align}
where we use the flat space derivative operator $d$. It can be extended to an appropriate covariant derivative (on e.g. a background instanton) if necessary.  The action is invariant under
\begin{align}\label{gauge-flat}
    \delta \omega^{AB}&= d\xi^{AB} +\mathcal{L}_\eta \omega^{AB}\,, &
    \delta \Psi^{ABCD}&= \mathcal{L}_\eta \Psi^{ABCD}\,,
\end{align}
where $\eta\equiv \eta^\mm(x)$ is a vector field. Modulo a gauge transformation, the effect of the $\eta$-symmetry on the one-forms $\omega^{AB}$ can be massaged into the following simpler form, cf. footnote \ref{ft:Cartan},
\begin{align}
    \delta_\eta \omega^{AB}&= i_\eta  (d\omega^{AB})\,,&
    \delta \Psi^{ABCD}&= i_\eta d\Psi^{ABCD}\,.
\end{align}
The action only depends on the exterior derivative of one-forms $\omega^{AB}$ and is thus invariant under shifts of these by an exact form. It is also clearly diffeomorphism invariant. This demonstrates that (\ref{gauge-flat}) are indeed the gauge symmetries. 

Now, we expand the action over the flat space background, where an appropriate solution for $\omega_0^{AB}$ is $\omega^{AB}_0=x\fud{A}{C'}dx^{BC'}$ in Cartesian coordinates (recall that $\omega^{AB}$ is not a spin-connection), so that $H^{AB}\equiv e\fud{A}{C'}\wedge e^{BC'}=\nabla \omega_0^{AB}$. One finds \begin{align}
    S_{\text{SDGRA, flat}}&= \int \Psi^{ABCD} \wedge H_{AB} \wedge d\omega_{CD} +\tfrac12  \int \Psi^{ABCD} \wedge d\omega_{AB} \wedge d\omega_{CD} \,,
\end{align}
where the first term is the correct free action \eqref{niceaction}. 
Upon imposing the light-cone gauge and introducing a coupling constant $g$ we arrive at the Siegel light-cone action for SDGRA \cite{Siegel:1992wd,Chalmers:1996rq}
\begin{align}\label{sdgr-lighcone}
    \int \Phi_{-2} \square \Phi_{+2}+g\, \epsilon_{A'B'}\epsilon_{C'D'} \Phi_{-2}\pl^{+A'}\pl^{+C'}\Phi_{+2}\, \pl^{+B'}\pl^{+D'}\Phi_{+2}\,.
\end{align}

\subsection{Higher spin generalisation: Flat space}
\label{sec:HS-flat}
We begin with flat space. The fields are represented by generating functions $\Psi$ and $\omega$. $\nabla$ is a covariant derivative such that $\nabla^2\omega\equiv 0$, i.e. its self-dual part vanishes and we can choose, for example, $\nabla\equiv d$. The action is simply 
\begin{align}\label{HSSDGRAflat}
    S&= \tfrac12 \action{\Psi}{F\wedge F }\,, & F&=\nabla \omega\,,
\end{align}
and is invariant under 
\begin{align}\label{flat-HS-symmetry}
    \delta \omega&= \nabla\xi + \overrightarrow{i_\eta} F\,, &\delta \Psi&= (  d \Psi) \overleftarrow{i_\eta}\,,
\end{align}
where for any $q$-form $X(y)\equiv X(y;x;dx)$ we define
\begin{align}\label{innerRL}
    X(y)\overleftarrow{i_\eta} &= X(y^A-\pl^A_1)\frac{\overleftarrow{\pl}}{\pl (dx^\mm)} \eta^\mm (y_1)\Big|_{y_1=0}\,, & \overrightarrow{i_\eta} X(y)&= \eta^\mm (y) \frac{\overrightarrow{\pl}}{\pl (dx^\mm)} X\,.
\end{align}
In the first definition $\pl^1_A\equiv \pl/\pl y_1^A$ will contract all spinor indices of $\eta$ with available indices of $X$. In the component form we find
\begin{align}\label{innerRA}
    X(y)\overleftarrow{i_\eta} &= \sum_{n,k}\tfrac{1}{n!} X\fdu{A(n)}{B(k)}\,\frac{\overleftarrow{\pl}}{\pl (dx^\mm)}\, \eta\fud{\mm;}{ B(k)} \, y^{A_1}...\,y^{A_n}\,.
\end{align}
To analyze the content of the theory we expand it over $\omega^{AB}_0=x\fud{A}{C'}dx^{BC'}$ and find
\begin{align}\label{action-HSSDGR-flat}
\begin{aligned}
    S&= \sum_{s=1} \tfrac{1}{(2s)!}\int \Psi^{A(2s)}\wedge H_{AA}\wedge  d \omega_{A(2s-2)}+ \\
    &+\sum_{s_1, s_2=1} \tfrac{1}{2(2s_1+2s_2-4)!}\int \Psi^{A(2s_1+2s_2-4)}\wedge d\omega_{A(2s_1-2)} \wedge d\omega_{A(2s_2-2)}\,,
\end{aligned}
\end{align}
where we recall that $H^{AB}\equiv e\fud{A}{C'}\wedge e^{BC'}=\nabla \omega_0^{AB}$. 

\subsubsection{Amplitudes}

We can now characterise the interactions appearing in the second line of (\ref{action-HSSDGR-flat}) by projecting them onto helicity states. We have already described the polarisation spinors for the free theory in \ref{subsec:ampl}. It remains to insert these states into the cubic vertex. 

On negative helicity states $\epsilon^-(k)$ the 2-forms $d \omega_{A(2s-1)}$ have only the ASD parts, because their SD parts vanish in view of the equation satisfied by these states. So, the only non-vanishing part of $d \omega_{A(2s-1)}$ is the spinor
\be
M^{s-1} \frac{q_{A_1} \ldots q_{A_{2s-2}}}{ (qk)^{(2s-2)}} k_{A'} k_{B'}\,.
\ee
We then obtain for $d\omega\wedge d\omega$ the contraction of two such spinors in their primed indices, which gives
\be
M^{s_1+s_2-2} [12]^2 \frac{q_{A_1} \ldots q_{A_{2s_1-2}}q_{B_1} \ldots q_{B_{2s_2-2}}}{ (q1)^{(2s_1-2)}(q2)^{(2s_2-2)}}\,.
\ee
We now contract this with a negative helicity spinor. The unprimed indices should match, which gives the condition $2s_3=2s_1+2s_2-4$. This means that there is precisely a factor of $1/M$ remaining in the amplitudes for all spins. The amplitude is then given by
\be
{\cal A}^{--+} = \frac{1}{M} [12]^2 \frac{(q3)^{(2s_1+2s_2-4)}}{ (q1)^{(2s_1-2)}(q2)^{(2s_2-2)}} \delta(s_3-(s_1+s_2-2))\,.
\ee
Eliminating factors of the auxiliary spinor $q$ using the momentum conservation we get
\be
{\cal A}^{--+} = \frac{1}{M} \frac{[12]^{(2s_1+2s_2-2)}}{ [23]^{(2s_1-2)}[13]^{(2s_2-2)}} \delta(s_3-(s_1+s_2-2)).
\ee
At $s=2$ this is the usual result in gravity. 
\subsubsection{Light-cone gauge}
In the light-cone gauge the new higher spin action features minimal gravitational interactions and reads
\begin{align}\label{flatHSSDGRA}
    S&=\sum_s\int \Phi_{-s} \square \Phi_{+s}+g\sum_{-s_1+s_2+s_3=2}a_{s_1,s_2} \epsilon_{A'B'}\epsilon_{C'D'} \Phi_{-{s_1}}\pl^{+A'}\pl^{+C'}\Phi_{+s_2}\, \pl^{+B'}\pl^{+D'}\Phi_{+s_3}
\end{align}
When restricted to helicities $\pm2$ the action reproduces the light-cone action \eqref{sdgr-lighcone} of SDGRA proposed by Siegel \cite{Siegel:1992wd}. Action \eqref{flatHSSDGRA} corresponds to a contraction of the Chiral HiSGRA that was considered in \cite{Ponomarev:2017nrr}. In the language of \cite{Ponomarev:2017nrr} it corresponds to the Poisson contraction of the (kinematic) algebra. Again, $a_{s_1,s_2}$ are specific coefficients coming from the covariant action. We discuss the relations between this and other results in the literature in section \ref{sec:discussion}. For now, let us just note that it is this two-derivative gravitational interaction of higher spin fields in flat space that for a long time was thought not to exist  within the Fronsdal approach. At the same time, this interaction is clearly seen in the light-cone gauge and in the spinor-helicity language. For the first time we have a covariant description of this interaction. 

\subsection{Higher spin generalisation: Anti-de Sitter space}
\label{sec:}
We would like to combine the higher spin action \eqref{HSSDGRAflat} from the previous section with the SDGRA action \eqref{SDGRAKirill}, so that the resulting theory has constant (non-vanishing) curvature space as a natural vacuum. The fields are the same: zero-form $\Psi\equiv \Psi(y|x)$ and one-form $\omega\equiv \omega(y|x)$. A new ingredient that we will need is the Poisson bracket on the $C[y]$-space. Given any two functions $f(y)$ and $g(y)$ we define 
\begin{align}
    \{f,g\}&= \pl^C f \pl_Cg= \sum_{n,m} \tfrac{1}{(n-1)!(m-1)!} f\fdu{A(n-1)}{C}g_{A(m-1)C}\, y^A...\,y^A\,.
\end{align}
Thus, the Poisson bracket of objects $f,g$ involves the contraction of a single pair of spinor indices of $f^{A(n)}, g^{A(m)}$. With the help of the Poisson bracket we define
\begin{align}
    F&= d\omega+\tfrac12\{\omega,\omega \}\,, & \delta \omega&=d\xi +\{\omega,\xi\}\equiv D\xi\,.
\end{align}
For spin two this reproduces the formulas familiar from SDGR. Indeed, in this case $\omega=\tfrac12 \omega_{AA} y^A y^A$ and $\xi=\tfrac12 \xi_{AA} y^A y^A$. Let us also note that for $\omega$ of a general spin and the gauge parameter $\xi$ that has only two spinor indices and is thus a Lie algebra $sl_2$ element, the Poisson bracket $\{\omega,\xi\}$ leads to the canonical action of $sl_2$ on spin-tensors. 

We postulate the action to have the same form as \eqref{HSSDGRAflat}
\begin{align}\label{HSSDGRAads}
   S&=\tfrac12\action{\Psi}{F\wedge F} = \sum_{n,m=0}\frac{1}{2(n+m)!}\int \Psi^{A(n+m)}\wedge F_{A(n)}\wedge F_{A(m)}\,.
\end{align}
The main task is now to show that this action is gauge invariant under some non-linear extension of \eqref{flat-HS-symmetry}. 

Let analyze the $\xi$-variation of $\omega$ first. Using $\delta F=\{F,\xi\}$ we have
\begin{align}\label{stuckA}
     \delta_\xi S&=\tfrac12\action{\Psi}{\{F, \xi \}\wedge F}+\tfrac12\action{\Psi}{ F \wedge \{F ,\xi \}}=\tfrac12\action{\Psi}{\{F\wedge F, \xi \}}=-\tfrac12\action{\Psi}{\{\xi, F\wedge F \}}
\end{align}
We now need to define the action of $\xi$ on $\Psi$ so as to cancel this arising variation. And indeed, we observe that one can define a sensible action of the Poisson algebra on $\Psi$. First, for any $f$, $g$ and $\xi$ in $C[y]\otimes \Omega(M)$ we define (the fact that $\action{\bullet}{\bullet}$ involves integration is not important at the moment and we can work with $\scalar{\bullet}{\bullet}$, which is just the full contraction of $sl_2$-indices, \eqref{scalarp})
\begin{align}
     \scalar{ f}{ \{\xi,g\}}&:=  \scalar{ f\circ\xi}{g}\,.
\end{align}
Thus, the idea is to define a new $\circ$ action that is the adjoint with respect to $\scalar{\bullet}{\bullet}$ of the action by the Poisson bracket. A simple computation shows that 
\begin{align}
    \Psi\circ \xi&= \sum \tfrac{1}{m!}\Psi\fud{B(n)}{A(m)}\xi_{B(n)A}\, y^A...y^A \,.
\end{align}
We draw reader's attention to the fact that all but one indices of $\xi$ have to be contracted with $\Psi$. 

It is now convenient to encode various operations on $C[y]$ in terms of generating functions. Let us define $(p_i)_C\equiv \pl^{y_i}_C$ and $a\cdot b\equiv a^C b_C$. Then we have
\besubeqs
\begin{align}
     \{f,g\}&= (p_1 \cdot p_2)\exp[y\cdot(p_1+p_2)] f(y_1)g(y_2)\Big|_{y_{1,2}=0}\,,\\
     f\circ \xi&= (y\cdot p_2) \exp[y\cdot p_1 + p_1\cdot p_2] f(y_1) \xi(y_2)\Big|_{y_{1,2}=0}\,,\\
     \left\langle f, g\right\rangle&=\exp[p_1 \cdot p_2] f(y_1)g(y_2)\Big|_{y_{1,2}=0}\,.
\end{align}
\esubeqs
The $\circ$-operation equips us with the (right) action of the Poisson algebra on the dual space:
\begin{align}
    \mathcal{R}_f (\Psi)&:=-\Psi\circ f\,, && [\mathcal{R}_f,\mathcal{R}_g] (\Psi)= \mathcal{R}_{\{f,g\}} (\Psi) \,.
\end{align}
We now continue with \eqref{stuckA} by adding the yet to be determined $\xi$-variation of $\Psi$ 
\begin{align}
     \delta_\xi S&=-\tfrac12\action{\Psi\circ \xi}{F\wedge F}+\tfrac12\action{\delta_\xi\Psi}{F\wedge F}\,.
\end{align}
Therefore, we need to postulate
\begin{align}
    \delta_\xi \Psi&= \Psi\circ \xi 
\end{align}
to make the action $\xi$-invariant. Note, that the $\circ$-action reduces to the standard $sl_2$ transformations for $\xi=\tfrac12 \xi_{AA}\,y^A y^A$. This shows that the action is invariant under "gauge" transformations generated by objects $\xi$ with an aritrary number of spinor indices. 

Let us now analyse the "shift" $\eta$-symmetry. We first postulate that its action on the connection $\omega$ is given by the insertion of a $y$-dependent vector field $\eta^\mu(y)$ into the curvature
\begin{align}\label{eta-action}
    \delta_\eta \omega&:= \overrightarrow{i_\eta}\, F\,, & \overrightarrow{i_\eta}F&= \eta^\mm(y) \frac{\overrightarrow{\pl}}{\pl dx^\mm} F(y;dx)\,.
\end{align}
Let us vary the action with respect to $\omega$:
\begin{align}
\begin{aligned}\label{var-omega}
    \delta_\omega S&= \tfrac12\delta_\omega \action{ \Psi} {F\wedge F}=\tfrac12\action{\Psi}{ D\delta \omega\wedge F+F\wedge D\delta \omega}=\action{ \Psi}{  D(\delta \omega\wedge F)}=\\
    &= \action{ \Psi}{ d(\delta \omega\wedge F)+\{\omega,\delta \omega\wedge F\} }= -\action{ d\Psi - \Psi\circ\omega}{  \delta \omega\wedge F}\,.
\end{aligned}
\end{align}
It makes sense to define the covariant derivative in the module as
\begin{align}
    R:= D\Psi = d\Psi - \Psi\circ\omega\,.
\end{align}
For the gravitational background ($\omega$ is bilinear in $y$) the last term coincides with the usual action of the spin connection on a spin-tensor. Also, $R$ has canonical transformation properties $\delta R=R\circ \xi$ and obeys Bianchi identity $dR+R\circ \omega+\Psi\circ F\equiv0$. We now come back to the $\eta$-symmetry. Using (\ref{eta-action}), (\ref{var-omega}) we have:
\begin{align}
    \delta_\eta S&= -\action{ R}{  \overrightarrow{i_\eta} F  \wedge F}+\tfrac12\action{\delta_\eta \Psi}{F\wedge F}.
\end{align}
We then use $\overrightarrow{i_\eta} F  \wedge F=\tfrac12 \overrightarrow{i_\eta} (F  \wedge F)$. Finally, we define 
\begin{align}
   \action{ R\overleftarrow{i_\eta}}{ X}&:= \action{ R }{ \overrightarrow{i_\eta} X}, 
\end{align}
which should be compared to \eqref{innerRL}. Explicitly, in the language of generating functions 
\begin{align}
    R \overleftarrow{i_\eta}&= \exp[y\cdot p_1 +p_1\cdot p_2] R(y_1;dx) \frac{\overleftarrow{\pl}}{\pl dx^\mm} \eta^\mu(y_2)\Big|_{y_i=0}\,.
\end{align}
It is now clear that we should require that $\delta_\eta \Psi=R\overleftarrow{i_\eta}$. Summarizing, we have the following gauge transformation rules
\besubeqs
\begin{align}
    \delta \omega&=d\xi +\{\omega,\xi\}+ \overrightarrow{i_\eta}F\,, &  F&=d\omega+\tfrac12\{\omega,\omega \}\,,\\
    \delta \Psi&= \Psi\circ \xi +R\overleftarrow{i_\eta}\,, &R &= d\Psi - \Psi\circ\omega\,.
\end{align}
\esubeqs
Once these transformation rules are expanded over the constant curvature background we find, in the leading order, the expected
\begin{align}
    \delta \Psi&=0+... & \delta \omega&= \nabla \xi + i_\eta F_0+...\,,
\end{align}
where $F_0=\tfrac12 H_{AA}y^Ay^A$. The last term is the desired shift symmetry, cf. \eqref{tworep}.

It is worth making few comments about the action. The action features quartic and even quintic terms.\footnote{There is not much literature on interactions of higher spin fields beyond cubic order, see e.g. \cite{Dempster:2012vw,Bekaert:2015tva,Buchbinder:2015apa,Joung:2019wbl,Karapetyan:2021wdc,Fredenhagen:2019lsz}.} However, the quintic terms will vanish in the light cone gauge. It is important to stress that the $\eta$-symmetry is crucial for preserving the right number of physical degrees of freedom. In the free approximation this symmetry can be written as \eqref{tworep}
\begin{align}
    \delta_\eta \omega^{A(2s-2)}&= e\fud{A}{C'} \eta^{A(2s-3),C'}\,.
\end{align}
However, vierbein $e^{AA'}$ is not a field that is present in the action. Instead, it is defined by the background. In the interacting theory this field becomes dynamical, and not easily constructible from the basic fields. Therefore, the Noether procedure would face some difficulties when trying to go beyond cubic vertices. 

Another remark is that, while in the spin two case the $\eta$-symmetry can be interpreted in terms of diffeomorphisms, this is no longer the case in the higher spin case. Indeed, the Cartan formula does not help to provide $\eta$-symmetry with a more geometrical interpretation since
\begin{align}
    \mathcal{L}_\eta\omega &=di_\eta \omega+i_\eta d\omega=D(i_\eta \omega)+ i_\eta F-\frac12 i_\eta\{\omega,\omega\} -\{\omega, i_\eta \omega\}\,.
\end{align}
The last two terms do not cancel each other since the Poisson bracket acts on $\eta$ too ($\eta$ depends on $y$ for $s>2$). Therefore, the shift symmetry should be defined the way we did via $i_\eta F$ rather than $\mathcal{L}_\eta \omega$. This subtlety does not make any difference in the case of spin two, but for higher spins the only available option is the one with $i_\eta F$.

Let us finally remark on the difference between the two actions \eqref{HSSDGRAflat} and \eqref{HSSDGRAads}. One is suitable for an expansion over the flat space and the other one has $AdS_4$ as a natural background.  The difference between the two action is in $\{\omega,\omega\}$-terms, which can be introduced with some coupling constant to be sent to zero in the flat space limit. Therefore, the flat space limit is smooth. In more detail, we replace $\omega$ with $g\,\omega $, where $g=\sqrt{|\Lambda|}$ and $\Lambda$ is the cosmological constant, which leads to $g^{-1}F=d\omega +\tfrac{g}2\{\omega,\omega\}$ and we should add $g^{-2}$ in front of \eqref{HSSDGRAads} to ensure the smooth $g\rightarrow0$ limit.

\section{Discussion}
\label{sec:discussion}
We have constructed actions for massless (propagating) higher spin fields that feature gauge (one-derivative) and gravitational (two-derivative) interactions.\footnote{It is convenient to define 'the number of derivatives' as the number of transverse derivatives in the light-cone gauge (provided trivial interactions that are proportional to the Hamiltonian are eliminated). Such definition would be independent of how the physical degrees of freedom are embedded into one or another covariant formulation. In anti-de Sitter space one should refer to the maximal number of derivatives since there is a lower derivative tail. In any case, for a triplet of helicities $\lambda_i$ the number of derivatives so defined is $|\lambda_1+\lambda_2+\lambda_3|$, \cite{Bengtsson:1986kh,Metsaev:2018xip}. } There are two types of theories: higher spin extensions of SDYM and of SDGRA. The HS-SDGRA theory admits two formulations, one having flat space and another having anti-de Sitter space as natural backgrounds. The flat limit is smooth and it agrees, upon going into the light-cone gauge, with the already known contractions of the Chiral HiSGRA \cite{Ponomarev:2017nrr}. When expanded over the $AdS_4$-background the actions should agree with the classification of vertices in the light-cone gauge obtained in \cite{Metsaev:2018xip} and with contractions of the cubic terms of Chiral HiSGRA in $AdS_4$ \cite{Skvortsov:2018uru}. The actions we proposed for HS-SDYM and HS-SDGRA are complete, i.e. are gauge-invariant up to all orders,\footnote{In principle, it is easy to construct something nonlinear and gauge invariant by taking powers of the linearized gauge-invariant curvatures, e.g. powers of the linearized field-strength $F_{\mu\nu}$, linearized Riemann tensor $R_{\mu\nu,\lambda\rho}$ and higher spin generalizations thereof. However, this usually does not lead to interesting interactions. Here, we consider some of the most important interactions: gauge and gravitational.} which seem to be the first Lorentz-covariant examples of this kind. 

The actions described in this paper are based on a new formulation of massless higher spin particles on constant curvature and, more generally, (anti)self-dual backgrounds. As we reviewed, the fields $\Psi^{A(2s)}$ and $\omega^{A(2s-2)}$ (or gauge potential $\Phi^{A(2s-1),A'}$) naturally emerge in twistor theory. The same fields appear in the "connection" description of gravity \cite{Krasnov:2012pd}. The actions we propose generalise to higher spins the familiar Chalmers-Siegel \cite{Chalmers:1996rq} covariant action for SDYM, and the "connection" formalism action \cite{Krasnov:2016emc} of SDGRA. As is the case with any chiral formalism, positive and negative helicities are treated in a different way. It would be interesting to construct other vertices that appear in Chiral HiSGRA, and, perhaps, redo the classification of vertices in the approach to higher spins described here.  

In the context of AdS/CFT correspondence SDYM, SDGRA and higher spin generalizations thereof, HS-SDYM and HS-SDGRA, may provide interesting cases of duality. These examples are quantum consistent theories in anti-de Sitter space, some of which contain dynamical graviton. It would be interesting to compute AdS/CFT correlation functions in these theories and investigate CFT duals thereof. The corresponding correlation functions should be a subset of those in Chern-Simons vector models \cite{Skvortsov:2018uru}. The recently developed spinor-helicity techniques in $AdS_4$ \cite{Nagaraj:2020sji,Nagaraj:2019zmk,Nagaraj:2018nxq} should be an appropriate starting point.

It remains to discuss the puzzle that the gauge and gravitational interactions of higher spin fields have long been thought not to exist, see e.g. \cite{Aragone:1979hx, Vasiliev:1986bq, Boulanger:2006gr, Conde:2016izb},  or require some non-minimal vertices and anti-de Sitter space \cite{Vasiliev:1986bq, Zinoviev:2008jz,Bekaert:2010hw}. On the other hand, they are fine in the light-cone gauge \cite{Bengtsson:1986kh,Metsaev:1991mt,Metsaev:1991nb} and the spinor-helicity formalism \cite{Conde:2016izb}. We discuss these puzzles below. 

Let us briefly recall some of the important milestones in the study of higher spin interactions. We stick to the results obtained within the field theory approach, putting aside any $S$-matrix results, e.g. Weinberg low energy theorem \cite{Weinberg:1964ew}. (A) In \cite{Aragone:1979hx} it was argued that it is impossible to put higher spin fields on nontrivial gravitational backgrounds. (B) In \cite{Berends:1984rq,Boulanger:2006gr,Zinoviev:2008ck,Manvelyan:2010je} it was noted that higher spin interactions require more derivatives, in particular, there is no standard, i.e. having two derivatives, gravitational interaction for higher spin fields. Instead, the simplest $s-s-2$ interaction\footnote{A cubic vertex of spin $s_i$ fields is called $s_1-s_2-s_3$ interaction. By standard or minimal gravitational interactions we mean something like $\sqrt{g}\, T_{\mu\nu}g^{\mu\nu}$, where $T_{\mu\nu}$ is a stress-tensor, which has two derivatives for bosonic fields. Such coupling would induce diffeomorphism symmetries on the matter fields. } has more than two derivatives. (C) in \cite{Vasiliev:1986bq} a 'gravitational' interaction (and some other cubic interactions as well) of massless higher spin fields was constructed in $AdS_4$ up to the leading (cubic) level and it was argued to have a singular flat space limit. The 'gravitational' interaction of \cite{Vasiliev:1986bq} does not have the minimal form and features terms with two and more derivatives. (D) However, around the same time a complete classification of cubic interactions of helicity fields was obtained in the light-cone gauge \cite{Bengtsson:1983pd,Bengtsson:1986kh} in flat space. This list clearly contains a two-derivative gravitational interaction $s-s-2$ of higher spin fields. The same result is easy to see in the spinor-helicity formalism \cite{Benincasa:2007xk,Conde:2016vxs}. (E) Recently, a complete classification of cubic interactions in $AdS_4$ was obtained in the light-cone gauge \cite{Metsaev:2018xip} and in the spinor-helicity formalism \cite{Nagaraj:2018nxq}, and is in one-to-one with that in flat space. (F) Such low-derivative interactions, which are not reachable within the Fronsdal approach, are necessarily present in Chiral HiSGRA \cite{Metsaev:1991mt,Metsaev:1991nb,Ponomarev:2016lrm,Skvortsov:2018jea}, which is the minimal extension of gravity with higher spin fields (provided they also feature at least one genuine higher spin interaction).

There is an apparent tension between some of the results reviewed above. Firstly, (A,B,C) are obtained in the Fronsdal approach (A,B) or are equivalent to those (C), while (E,D) are about objective reality. In other words, non-existence or singularity of certain structures observed in (A,B,C) may represent a feature of a given covariant approach, rather than an actual fact about interactions of higher spin fields. Secondly, there is a perfect (D+E) agreement between cubic interactions in flat and anti-de Sitter spaces. The latter proves that (A,B,C) reveal some subtleties of the Fronsdal formulation rather than of the structure of higher spin interactions themselves. The subtleties are of two types: (I) certain interactions (low derivative ones) are invisible in the Fronsdal approach; (II) there are certain discrepancies between results in flat space and $AdS$.

Thus, one important feature of any fundamental approach (e.g. light-cone and S-matrix) is the presence of low derivative interactions \cite{Bengtsson:1986kh,Metsaev:1991nb,Metsaev:1991mt,Metsaev:1993ap} that are not seen \cite{Bengtsson:2014qza,Conde:2016izb} in the Fronsdal formulation,\footnote{The same result, i.e missing interactions, can be obtained by extrapolating the light-cone results from $d>4$ to $d=4$ \cite{Metsaev:2005ar}. The vertices in the light-cone gauge agree with the ones in terms of Fronsdal fields in $d>4$. } see e.g. \cite{Boulanger:2008tg,Bekaert:2010hp}. For example, it was shown \cite{Conde:2016izb, Sleight:2016xqq} that the two-derivative gravitational interactions of higher spin fields cannot be (at least locally) written in terms of Fronsdal fields. This paper resolves all these puzzles by constructing a covariant formulation of higher spin fields that features the minimal gravitational interactions (it should also allow one to write down the other low derivative interactions that are invisible in the Fronsdal approach).

In more detail, consider the gravitational interactions $s-s-2$ as an example. The light-cone offers vertices of type $(+s,-s,+2)$ (two derivatives),\footnote{Curiously enough \cite{Boulanger:2006gr} found the deformation of the gauge algebra and gauge transformations that would correspond to such a vertex in the Fronsdal language. However, it was shown there that they fail to give a consistent vertex.} $(+s,+s,-2)$ ($2s-2$ derivatives) and $(+s,+s,+2)$ ($2s+2$ derivatives, but we will not need this one). In flat space each vertex is one term with a fixed number of derivatives. When deformed to $AdS$ it has to be accompanied with a tail of lower derivative terms to remain consistent \cite{Metsaev:2018xip}. In $AdS_4$ the $(+s,+s,-2)$-vertex begins with a $(2s-2)$-derivative term and goes down to a $1$-derivative one. The $(+s,-s,+2)$-vertex consist of a two derivative term and a one-derivative term (tail).

What is observed in Fronsdal language? (B) can see only $(+s,+s,-2)$-vertex in flat space. (C) has $(+s,+s,-2)$-vertex together with its lower derivative completion due to $AdS_4$, see \cite{Boulanger:2008tg,Zinoviev:2008ck,Joung:2013nma}. The last terms in the tail have two and one derivatives and can be mixed up with the $(+s,-s,+2)$-vertex. (F) implies that $(+s,-s,+2)$-vertex must be present too. Therefore, our conclusion is that the covariant 'gravitational' vertex in $AdS_4$ that starts with $(2s-2)$ derivatives is a chimera: upon going into the light-cone gauge the vertex has to decompose into two independent vertices $(+s,+s,-2)$ and $(+s,-s,2)$ with a fixed relative coefficient. The coefficient can be read off from Chiral HiSGRA. The same has to be true for other low derivative interactions. It would be interesting to check this. We believe that this resolves (I).

As for (II), there are different ways to define flat limit:\footnote{For example, the 'gravitational' interaction in the Fronsdal approach would look schematically as $l_p(\Phi_2\nabla^2 \Phi_s\Phi_s+... + R^{2s-4}\Phi_2\nabla^{2s-2} \Phi_s\Phi_s)$, where $l_p$ is Planck length and $R$ is $AdS$ 'radius'.} (i) one can simply send the radius $R$ to $\infty$ as in \cite{Vasiliev:1986bq}, which is singular; (ii) one can rescale both $l_p$ and $R$ in such a way that the highest derivative term survives \cite{Bekaert:2005jf,Boulanger:2008tg,Zinoviev:2008ck}. This trick works for each vertex separately; (iii) one can assume that fluctuations take place far from the boundary $z=0$ and are of typical size $l$ that is much smaller than $R$, which would bring the entire Chiral HiSGRA from $AdS_4$ to flat space and make the flat limit smooth for all interaction vertices together.\footnote{Note that the vertices in the light-cone gauge in $AdS_4$ do not have any explicit dependence on $R$ at all \cite{Metsaev:2018xip}.} What confronting (A,B,C) with (D,E,F) shows is that the mixture of $(+s,+s,-2)$ and $(+s,-s,2)$ interactions that is present in the $s-s-2$ vertex in $AdS_4$ in the Fronsdal formulation boils down to a single $(+s,+s,-2)$ vertex in the flat limit. On the contrary the vertices we constructed so far have a smooth flat limit.

As the results of this paper and previous works suggest, there is no difference between flat space and $AdS$ for the problem of constructing interactions of massless higher spin fields. For example, the cubic interactions are in one-to-one. More generally, there should be no difference provided that all questions are formulated in an invariant way. Starting from the quartic order there are severe obstructions to existence of HiSGRA both in flat \cite{Bekaert:2010hp,Fotopoulos:2010ay,Ponomarev:2017nrr,Roiban:2017iqg} and AdS cases \cite{Bekaert:2015tva,Sleight:2017pcz,Ponomarev:2017qab}. Chiral HiSGRA (ant its contractions) is the only perturbatively local HiSGRA in flat space \cite{Ponomarev:2017nrr}, which is likely to be so in $AdS_4$. 

As a general lesson, it is convenient to stick to the approaches that deal directly with physical degrees of freedom (p.d.o.f.) and, thereby, avoid any ambiguities that are present in many local covariant field theory formulations (the same p.d.o.f. can be embedded into different covariant fields and some of these formulations may not be convenient for introducing interactions, see e.g. \cite{Bekaert:2002uh}). Two such invariant approaches come to one's mind: $S$-matrix and light-cone (or light-front) approach. The former idea is to try to construct a consistent $S$-matrix, but it does not come with too many tools. The latter idea is to build the charges of the Poincare (or any other spacetime's symmetry algebra), including the Hamiltonian in terms of local p.d.o.f. 

We prefer the light-cone approach as, perhaps, the most general approach to local field theory. Once the light-cone form of a theory is known, it makes a lot of sense to look for a covariant formulation, which would make it possible to deal with non-perturbative problems and simplify most of the light-cone computations even perturbatively. As far as we know, there is no theorem that would guarantee that a nice covariant formulation always exists. In this paper we make an observation that the most important interaction vertices that are present in Chiral HiSGRA and could not be written within the Fronsdal approach can now be written with the help of the new approach. The covariant fields that we employed were known since 70's thanks to the development of twistor theory. All this strongly suggests that Chiral HiSGRA should have a manifestly covariant action, which might be easier to look for directly in the twistor space. Other HiSGRA in $AdS_4$ can be understood as perturbations of the Chiral similarly to how YM and Gravity can be understood as deformations of SDYM and SDGRA.

\section*{Acknowledgments}
\label{sec:Aknowledgements}
Some of the results reported in this paper date back to 2012 and have been discussed with a number of colleagues over the years starting from "Recent Developments on Light Front" workshop during March 14-16, 2017 at LMU. We are indebted to Tim Adamo, Nicolas Boulanger, Maxim Grigoriev, Gregory Korchemsky, Yannick Herfray, Tristan McLoughlin, Ruslan Metsaev, Alexey Sharapov and especially to Karapet Mkrtchyan and Dmitry Ponomarev for useful discussions. The work of E.S. was supported by the Russian Science Foundation grant 18-72-10123 in association with the Lebedev Physical Institute. 

\footnotesize
\providecommand{\href}[2]{#2}\begingroup\raggedright\endgroup

\end{document}